\documentclass[aps,nofootinbib]{revtex4}

\usepackage{epsfig}

\newlength{\abstwidth}
\renewcommand{\d}{{\mathrm d}}

\def\be{\begin{equation}} 
\def\ee{\end{equation}}

\begin{document}

\def\lsim{\mathrel{\rlap{\lower4pt\hbox{\hskip1pt$\sim$}}
    \raise1pt\hbox{$<$}}}         
\def\gsim{\mathrel{\rlap{\lower4pt\hbox{\hskip1pt$\sim$}}
    \raise1pt\hbox{$>$}}}         

\pagestyle{empty}

\begin{flushright}
BI-TP 2002/05\\
\end{flushright}

\vspace{\fill}

\begin{center}
{\Large\bf Diffractive production and the total cross section in deep inelastic scattering$^*$}
\\[1.8ex]
{\bf Masaaki Kuroda} \\[1.2mm]
Institute of Physics, Meiji-Gakuin University \\[1.2mm]
Yokohama 244, Japan \\[1.2mm]
and \\[1.5ex]
{\bf Dieter Schildknecht} \\[1.2mm]  
Fakult\"{a}t f\"{u}r Physik, Universit\"{a}t Bielefeld \\[1.2mm] 
D-33501 Bielefeld, Germany \\[1.5ex]
\end{center}

\vspace{\fill}

\begin{center}
{\bf Abstract}\\[2ex]
\begin{minipage}{\abstwidth}
We explore the consequences for diffractive production, $\gamma^* p \rightarrow
X p$, in deep inelastic scattering at low values of $x \cong Q^2/W^2\ll 1$ that 
follow from our recent representation of the total photoabsorption cross 
section, $\sigma_{\gamma^* p}$, in the generalized vector dominance/color
dipole picture (GVD/CDP) that is based on the generic structure of the two-gluon 
exchange from QCD. Sum rules are derived that relate the transverse
and the longitudinal (virtual) photoabsorption cross section to diffractive
forward production of $q \bar q$-states that carry photon quantum numbers 
(``elastic diffraction''). 
Agreement with experiment in the $W^2$ and $Q^2$ 
dependence is found for $M^2_X/Q^2 \ll 1$, where
$M_X$ is the mass of the produced system $X$. An additional component (``inelastic
diffraction''), not 
actively 
contributing to the forward Compton amplitude, is needed for diffractive 
production at high values of $M_X$.
Our previous theoretical representation of the total photoabsorption 
cross section, 
$\sigma_{\gamma^* p} = \sigma_{\gamma^* p} (\eta)$, in terms of the scaling
variable $\eta \equiv (Q^2 + m^2_0) / \Lambda^2 (W^2)$ is extended to 
include the entire kinematic domain, $x \le 0.1$ 
and all $Q^2$ with $Q^2 \ge 0$, 
where scaling in $\eta$ holds experimentally. 
\end{minipage}
\end{center}

\vspace{\fill}
\noindent

\rule{60mm}{0.4mm}\vspace{0.1mm}

\noindent
${}^*$ Supported by the BMBF, Contract 05 HT9PBA2 and by the Ministry 
of Education, Science and Culture in Japan under the Grant-in-Aid 
No.14340081.\\
\clearpage
\pagestyle{plain}
\setcounter{page}{1}

\section{Introduction}

Any theory of diffractive production, $\gamma^* p \rightarrow Xp$ in deep 
inelastic scattering (DIS), at low $x = Q^2/W^2 \ll 1$ has to 
discriminate between, and take into account two distinctly different components, 
an ``elastic'' and an ``inelastic'' one. The elastic component, by 
definition, consists of all and only those final states $X$ that 
carry the quantum numbers of 
the photon. Accordingly, the (imaginary) elastic diffractive 
production amplitude
is responsible for the (virtual)
forward Compton scattering amplitude that, via the optical theorem, represents 
the total 
photoabsorption cross section, $\sigma_{\gamma^* p} (W^2, Q^2)$. The inelastic
component contains hadronic states $X$ that do not carry photon quantum numbers.
In particular, the inelastic component contains states $X$ of spins different 
from the spin of one unit carried by the photon. It, obviously, does not 
contribute to the Compton forward scattering amplitude. 

Direct evidence for an inelastic component in diffractive production in DIS is 
provided by the experimentally observed properties of the hadronic state in 
diffractive DIS. The thrust and sphericity distributions of the state $X$ are 
approximately identical \cite{ZEUS2} to the ones observed in 
$e^+ e^- \rightarrow q \bar q \rightarrow$ hadrons. 
The strong alignment of the jet axis of the 
$(q \bar q)$ state $X$ in its rest frame with respect to the virtual photon 
direction, however, is different from the $1 + \cos^2 \theta$ dependence 
observed in $e^+ e^-$ annihilation, and this difference provides evidence 
for the presence of 
higher spin states in the system $X$. Also, in hadron-hadron interactions, 
diffractive production of states, different in spin from the initial beam 
particles, is a common feature. The change in spin is connected with a change 
in parity that obeys the natural-spin-parity connection, $(-1)^{\Delta J}$,
\cite{Goul}. 

In the present investigation, we consider elastic diffraction in its 
connection with the total virtual photoabsorption cross section. In terms of the 
virtual forward Compton scattering amplitude, DIS at low $x$ corresponds to 
diffractive scattering of the hadronic $(q \bar q)$ states the photon 
fluctuates into, as conjectured by generalized vector dominance \cite{Sakurai}
a long time ago. The $q \bar q$ states the photon is coupled to interact 
via two-gluon exchange \cite{4}
with the target nucleon. The $q \bar q$ states form a color dipole, and the 
forward scattering amplitude resulting from two-gluon exchange becomes 
diagonal, when expressed in terms of the (transverse) quark-antiquark 
separation in position space \cite{6}. When transformed into momentum 
space, off-diagonal transitions of destructive nature appear with respect to 
the masses in the propagators of the incoming and outgoing $q \bar q$ states, 
essential for convergence of the whole formalism, as anticipated by off-diagonal
generalized vector dominance \cite{5} in the pre-QCD era\footnote{Compare also
ref.\cite{Frankfurt} for a discussion of the relevance of off-diagonal 
generalized vector dominance}. Our approach 
\cite{1,2,3} to DIS that is based on the generic two-gluon exchange 
structure and incorporates the empirical scaling of $\sigma_{\gamma^* p}
(W^2, Q^2) = \sigma_{\gamma^* p} (\eta (W^2 , Q^2))$ with $\eta \equiv 
(Q^2 + m^2_0) / \Lambda^2 (W^2)$, \cite{1}, is appropriately called the 
generalized vector dominance/color-dipole picture (GVD/CDP). 

In the present paper we will show that the total cross section, $\sigma_{
\gamma^* p}$, at small $x$ can explicitly be represented by a sum rule that
contains the elastic diffractive production amplitude integrated over the 
masses of the diffractively produced states $X$. The direct connection of 
$\sigma_{\gamma^* p}$ and diffractive production by the sum rule will be seen 
to suggest an improved expression for $\sigma_{\gamma^* p}
= \sigma_{\gamma^* p} (\eta)$ in the GVD/CDP, such that the GVD/CDP covers 
the full 
kinematic range where scaling in $\eta$ was established \cite{1}, 
i.e. $x < 0.1$ with all $Q^2 \ge 0$. 

A comparison of the results for elastic diffraction with the experimental 
data for $\gamma^* p \rightarrow Xp$ reveals the presence of a large excess
in high-mass production, i.e. for $\beta \equiv Q^2 / (Q^2 + M^2_X)\ll 1$.
This excess is to be associated with inelastic diffractive production. 

In Section 2, we briefly present the GVD/CDP for $\sigma_{\gamma^* p}$. In 
Section 3, we describe elastic diffraction. A comparison of the representations 
in Sections 2 and 3 implies the sum rules of Section 4. The comparison with 
the experimental data for diffractive production in Section 5 reveals 
approximate agreement for small diffractively produced masses, i.e. for 
$\beta \rightarrow 1$, while showing the mentioned excess for $\beta 
\rightarrow 0$. The improved representation of $\sigma_{\gamma^* p}(\eta)$ 
is given in Section 6, while Section 7 summarizes our main conclusions.

\section{The virtual forward Compton Scattering amplitude and $\sigma_{\gamma^* 
p}$}

The virtual forward Compton scattering amplitude at low $x$, as mentioned, 
results from the diffractive scattering of $q \bar q$ states on the proton,
\be
q \bar q + {\rm proton} \rightarrow q \bar q + {\rm proton} , 
\label{(1)}
\ee
where both the incoming and outgoing $q \bar q$ pair carry the  
quantum numbers of the photon. In particular, the incoming and outgoing 
$q \bar q$ states in 
(\ref{(1)}) have spin 1. 
The $x \rightarrow 0$ limit of the 
two-gluon-exchange virtual forward Compton scattering amplitude is 
embodied \cite{1} 
in the position-space representation \cite{6}
for the transverse and longitudinal parts of the photoabsorption cross section, 
$\sigma_{\gamma^*_T p}$ and $\sigma_{\gamma^*_L p} $, 
\be
\sigma_{\gamma^*_{T,L} p} (W^2 , Q^2) = 
\int dz \int d^2 r_\bot \sum_{\lambda, \lambda^\prime = \pm 1}
| \psi_{T,L}^{(\lambda , \lambda^\prime)} ( \vec r_
\bot , z, Q^2) |^2 
\sigma_{(q \bar q)p} (\vec r_\bot^{~2} , z, W^2)
\label{(2)}  
\ee
with the Fourier representation of the color dipole cross section, 
\be
\sigma_{(q \bar q)p} (\vec r_\bot^{~2} , z, W^2) = \int d^2 l_\bot 
\tilde\sigma_{
(q \bar q)p} (\vec l^{~2}_\bot , z, W^2)(1 - e^{-i\vec l_\bot \vec r_\bot})
\label{(3)}
\ee
that incorporates color transparency and (hadronic) unitarity \cite{1}. In
(\ref{(2)}) and (\ref{(3)}), we use the conventional notation, in which
$\vec r_\bot$ denotes the transverse interquark separation and $z$ the
fraction of the (virtual) photon momentum carried by the quark. The
transverse gluon momentum is denoted by $\vec l_\bot$.
The square of the photon wave function reads\footnote{In contrast to 
ref \cite{1}, we include the color factor $N_c = 3$ in the wave function 
squared.}
\begin{eqnarray}
\sum_{\lambda,\lambda^\prime}\Big| \psi_{T,L}^{(\lambda, \lambda^\prime)} 
(\vec r_\bot , z; Q^2) \Big|^2 = 
& & 3 \cdot \frac{4\pi}{(16\pi^3)^2} \int d^2 k^{~\prime}_\bot 
\int d^2
k_\bot {\cal M}^*_{T,L} (\vec k^{~\prime}_\bot , z , Q^2) 
\nonumber  \\        
& & {\cal M}_{T,L} (\vec k_\bot , z ,Q^2) 
\exp (i \vec k^{~\prime}_\bot - \vec k_\bot ) \vec r_\bot , 
 \label{(4)} 
\end{eqnarray}
where  
\be
{\cal M}^*_T (\vec k^{~\prime}_\bot , z, Q^2) \cdot {\cal M}_T (\vec k_\bot, 
z, Q^2) = \frac{8\pi \alpha (\vec k^{~\prime}_\bot \cdot \vec k_\bot )
\sum_f Q^2_f (z^2 + (1 - z)^2)}{(z (1-z)Q^2 + \vec k^{~
\prime 2}_\bot) (z (1-z) Q^2 + \vec k^{~2}_\bot)}
\label{(5)}
\ee
and
\be
{\cal M}^*_L (\vec k^{~\prime}_\bot , z, Q^2) \cdot {\cal M}_L (\vec k_\bot, 
z; Q^2) = \frac{32\pi \alpha Q^2 \sum_f Q^2_f z^2  (1 - z)^2}
{(z (1-z)Q^2 + \vec k^{~
\prime 2}_\bot )(z (1-z) Q^2 + \vec k^{~2}_\bot)}.
\label{(6)}
\ee
Substitution of the photon wave function (\ref{(4)}) and the dipole cross 
section (\ref{(3)}) into (\ref{(2)}) takes us back to the momentum space 
representation of the photoabsorption cross section  
\begin{eqnarray}
\sigma_{\gamma^*_{T,L} p} (W^2 , Q^2) = & & 
\frac{3}{16\pi^3 \cdot 2} \int dz  \int d^2 k_\bot
\int d^2 l_\bot \tilde\sigma_{(q \bar q)p}
(\vec l^{~2}_\bot , z, W^2)  \cdot \nonumber \\
& & \cdot | {\cal M}_{T,L} (z, \vec k_\bot , Q^2) - {\cal M}_{T,L} (z, \vec k_
\bot + \vec l_\bot , Q^2) |^2 . 
\label{(7)}
\end{eqnarray}
Here, $\vec k_\bot$ denotes the transverse momentum of the quark in the
$q \bar q$ state that originates from the (virtual) photon.
The two-gluon exchange interaction of the $q \bar q$ pair in (\ref{(7)}) 
involves
integration over the (transverse) momentum, $\vec l_\bot$, of the gluon.
Guided by the empirical scaling law\cite{1},
\be
\sigma_{\gamma^* p} (W^2, Q^2) = \sigma_{\gamma^*p} (\eta (W^2, Q^2)),
\label{(8)}
\ee
with
\be
\eta(W^2, Q^2) = \frac{Q^2 + m^2_0}{\Lambda^2 (W^2)},
\label{(8a)}
\ee
where $\Lambda^2 (W^2)$ increases slowly with energy and $m_0$ denotes
a threshold mass, the distribution in the gluon momentum in the GVD/CDP
is approximated\cite{1,2} by a $\delta$-function situated at the average 
(or effective)
gluon momentum determined by $\Lambda (W^2)$, i.e.
\be
\tilde \sigma_{(q \bar q)p} (\vec l^{~2}_\bot, z, W^2) = \sigma^{(\infty)} 
\frac{1}{\pi} \delta (\vec l^{~2}_\bot - z (1-z) \Lambda^2 (W^2)).
\label{(8b)}
\ee
The asymptotic value of the color dipole cross section (\ref{(3)}), 
$\sigma^{(\infty)}$, turned out to be constant in good approximation\cite{1}.
The factor $z(1-z)$ in (\ref{(8b)}) is a model assumption that is particularly 
relevant at $Q^2 \gg m^2_0$.
The energy dependence of 
$\Lambda^2 (W^2)$ was parameterized\cite{1}, alternatively, 
by a power law or by a 
logarithm, 
\be
\Lambda^2 (W^2) = \left\{ \matrix{ & C_1 (W^2 + W^2_0)^{C_2} , \cr
  & C^\prime_1 \ln \left( \frac{W^2}{W^2_0} + C^\prime_2 \right).\cr}
\right.
\label{(9)}
\ee

For later reference, we note the representation of the transverse and the 
longitudinal cross section (\ref{(7)}) obtained upon substitution of 
(\ref{(8b)}) and upon having carried out all integrations except the one over 
$\vec k^{~2}_\bot$. In terms of the $q \bar q$ mass, 
\be
M^2 = \frac{\vec k^{~2}_\bot}{z (1-z)},
\label{(10a)}
\ee
we find\footnote{Here we ignore the additive ``correction terms'' \cite{1} 
that 
assure an identical threshold mass, $m_0$, for the incoming and outgoing
$q \bar q$ pair in the forward Compton amplitude. The correction terms
will be given below.
For the transverse cross section, the correction is negligible, while 
in the longitudinal
one, contributions are of the order of 10 \%.}
\begin{eqnarray}
& & \sigma_{\gamma^*_T p} (W^2 , Q^2) = 
\frac{\alpha R_{e^+ e^-}}{3\pi} \sigma^{(\infty)}  \label{(11a)} \\ 
& & \cdot \int_{m^2_0} d M^2 \frac{1}{Q^2 + M^2} 
\left[ \frac{M^2}{Q^2 + M^2} - \frac{1}{2} \left( 1 + \frac{M^2 - \Lambda^2 
(W^2) - Q^2}{\sqrt{(M^2 + \Lambda^2 (W^2) + Q^2)^2 - 4 \Lambda^2 (W^2) M^2}} 
\right) \right]
\nonumber
\end{eqnarray}
and
\begin{eqnarray}
& & \sigma_{\gamma^*_L p} (W^2 , Q^2) =
\frac{\alpha R_{e^+ e^-}}{3\pi} \sigma^{(\infty)}  \label{(12a)} \\
& & \cdot \int_{m^2_0} d M^2 \frac{1}{Q^2 + M^2} 
\left[ \frac{Q^2}{Q^2 + M^2}- \frac{Q^2}{\sqrt{(M^2 + \Lambda^2 (W^2) + Q^2)^2 
- 4 \Lambda^2 (W^2) M^2}} 
\right]. \nonumber 
\end{eqnarray}
The total photoabsorption cross section, $\sigma_{\gamma^* p}
(W^2 , Q^2)= \sigma_{\gamma^*_T p} + \sigma_{\gamma^*_L p}$, 
scales in the variable $\eta$ given by (\ref{(8a)}), i.e.
\be
\sigma_{\gamma^* p} (W^2 , Q^2) = \frac{\alpha R_{e^+ e^-}}{3\pi} \,
\sigma^{(\infty)} I^{(1)} (\eta) ,
\label{(11)}
\ee
with
\begin{eqnarray}
& & I^{(1)} \left( \eta, \mu \equiv \frac{m^2_0}{\Lambda^2 (W^2)} \right)
\label{(11b)} \\
& & = \frac{1}{2} ln \frac{\eta - 1 - \sqrt{(1 + \eta)^2 - 4 \mu}}{2 \eta}
+ \frac{1}{2 \sqrt{1 + 4 (\eta - \mu)}} \nonumber \\
& & \times ln \frac{\eta (1 + \sqrt{1 + 4 (\eta - \mu)}}{4 \mu - 1 - 3 \eta 
+ \sqrt (1 + 4 (\eta - \mu))((1 + \eta)^2 - 4 \mu)}, \nonumber
\end{eqnarray}
the dependence of $I^{(1)}$ on $m^2_0/\Lambda^2(W^2)$ being negligible for
$m^2_0/\Lambda^2 (W^2) << 1$. We also note the asymptotic form
\be
\sigma_{\gamma^* p} (W^2 , Q^2) = \frac{\alpha R_{e^+ e^-}}{3\pi} 
\sigma^{(\infty)}
\left\{ \matrix{ \ln (1/\eta) , & {\rm for}\, \eta\rightarrow 
m^2_0/\Lambda^2 (W^2) , \cr
1 / 2\eta, & {\rm for} \, \eta \gg 1 . \cr } \right.
\label{(12)} 
\ee
The agreement of the photoabsorption cross section with photoproduction for 
$Q^2 \rightarrow 0$ determines the product  
\be
R_{e^+ e^-} \sigma^{(\infty)}= \sigma^{(\infty)} 3 \sum_f Q^2_f . 
\label{(13)} 
\ee
With three active quark flavors, $R_{e^+e^-} =2$, we have 
$\sigma^{(\infty)} \cong 80 {\rm GeV}^{-2} \cong 31$ mb, while with 
four active 
flavors, $R_{e^+e^-} = 10/3$ and $\sigma^{(\infty)} \cong 48 
{\rm GeV}^{-2} \cong 
18.7$ mb \cite{1}. For further reference, we note the parameters entering 
the scaling variable $\eta$ and 
the $W$ dependence of $\Lambda^2 (W^2)$, as determined by the fit to the
total cross section \cite{1}. For the power law in (\ref{(9)}) we have
\begin{eqnarray}
m^2_0 & = & 0.16 \pm 0.01 {\rm GeV}^2 ,~~~~~~~~~~~ W^2_0  =  
882 \pm 246 {\rm GeV} ,
\nonumber \\
C_1 &= & 0.34 \pm 0.05 ({\rm GeV}^2)^{1-C_2},~~~ C_2  =  0.27 \pm 0.01 , 
\label{(14)}  
\end{eqnarray}
while for the logarithmic dependence, 
\begin{eqnarray}
m^2_0 & = & 0.157 \pm 0.009 {\rm GeV}^2 ,~~~ W^{\prime 2}_0  =  1015 
\pm 334 {\rm GeV}^2, \nonumber \\
C_1^\prime &= & 1.644 \pm 0.14 {\rm GeV}^2,~~~~~ C_2^\prime  =  4.1 \pm 0.04.
\label{(15)}  
\end{eqnarray}
We note that the GVD/CDP with scaling in $\eta$ leads to the important conclusion
\cite{2} that 
\be
\lim_{{W^2 \rightarrow\infty}\atop{Q^2 {\rm fixed}}} \, \frac{\sigma_{\gamma^* p}
(W^2 , Q^2)}{\sigma_{\gamma p} (W^2)} = 1 , 
\label{(15a)}
\ee
i.e. virtual and real photons have the same cross section at infinite energy
(``saturation'').

\section{Elastic diffraction}

We turn to diffractive production. The diagonal form (\ref{(2)}) of $\sigma_{
\gamma^*_{T,L} p}$ in transverse position space develops its full power when 
considering diffractive (forward) production, $\gamma^* p \rightarrow
Xp$.  Indeed, the 
diffractive production  cross section of state $X$ of spin 1
in the forward direction, via the two-gluon exchange generic structure 
$(x\rightarrow 0)$ becomes \cite{7},
\begin{eqnarray}
& & \frac{d\sigma_{\gamma^*_{T,L}p\rightarrow X p}(W^2, Q^2, t)}{dt} 
\Bigg|_{t=0} \label{(16)} \\ 
& & = \frac{1}{16\pi} \int^1_0 dz
\int d^2 r_\bot \sum_{\lambda , \lambda^\prime = \pm 1} 
| \psi_{T,L}^{(\lambda , \lambda^\prime)}(r_\bot , z, Q^2) |^2 
\sigma_{(q \bar q)p}^2 (\vec r^{~2}_\bot , z, W^2) .
\nonumber
\end{eqnarray}
The representation
(\ref{(16)}) contains the square of (the imaginary part\footnote{Neglecting
 a potential contribution from a real part to the $(q\bar q)p$ forward 
scattering amplitude seems justified on phenomenological grounds from the
empirical knowledge on photon- and hadron-induced reactions \cite{Goul}.
For a brief theoretical estimate of the (small) ratio of the real to imaginary 
part of the $(q\bar q)p$ forward scattering amplitude with two-gluon-exchange
interaction, compare with ref.\cite{levin}.}
 of) the forward
production amplitude for reaction (1) that enters (2) linearly and 
necessarily only involves $q \bar q$ pairs that couple to the photon and 
accordingly carry photon quantum numbers (``elastic diffraction''). 
The factor $1/16\pi$ in (\ref{(16)}) stems from the application of the 
optical theorem 
when passing from the forward scattering amplitude of reaction 
(\ref{(1)}) to the total cross section, $\sigma_{(q \bar q)p}$.
Diffractive production of higher spin states (inelastic diffraction)
requires an
additive term, $\sigma^2_{(q \bar q) p} \rightarrow \sigma^2_{(q \bar q)p}
+ \Delta \sigma^2_{(q \bar q)p}$ in (\ref{(16)}).

Note the close analogy of (\ref{(16)}) to the simple $\rho^0$  
dominance formula, where in photoproduction \cite{7a}
\be
\frac{d\sigma}{dt} \Bigg|_{t=0} (\gamma p \rightarrow \rho^0 p) = \frac{1}
{16\pi} \frac{\alpha \pi}{\gamma^2_\rho} \sigma^2_{\rho^0 p} . 
\label{(17)}
\ee 
The constant $\alpha \pi/\gamma^2_\rho$ denotes the photon-$\rho^0$ coupling
strength as measured in $e^+ e^-$-annihilation. 
The generalization (\ref{(16)}) of (\ref{(17)}) is an outgrowth of 
the diagonalization of the 
process $\gamma^* p \rightarrow X p$
that is specific to the use of the variables $\vec r_\bot$ and $z$. 
It is precisely with respect to these variables that the process $\gamma^*
p \rightarrow Xp$ is truly elastic: a $q \bar q$ dipole being specified by 
$\vec r_\bot$ and $z$ and carrying photon quantum numbers undergoes elastic 
forward scattering. 

Inserting the photon wave function (\ref{(4)}) as well as the 
representation (\ref{(3)}) for the dipole cross section into (\ref{(16)}), we 
obtain the momentum space representation 
\begin{eqnarray}
& & \frac{d\sigma_{\gamma^*_{T,L}p\rightarrow X p}(W^2, Q^2, t)}{dt} 
\Bigg|_{t=0} =  \frac{1}{16\pi} \frac{3}{16\pi^3} 
\int^1_0 dz \int d^2 k_\bot \cdot \label{(18)}  \\
& & \cdot \left[ \int d^2 l_\bot \tilde\sigma_{(q \bar q)p} 
(\vec l^{~2}_\bot , z, W^2 ) ( 
{\cal M}_{T,L} (\vec k_\bot , z , Q^2) - {\cal M}_{T,L} (\vec k_\bot + 
\vec l_\bot , z , Q^2 ) ) \right]^2 . \nonumber
\end{eqnarray}
We emphasize the destinctive difference between (\ref{(7)}) and (\ref{(18)}) 
with respect to the order in which the square of the integrand is taken 
and the  
integration over the transverse gluon momentum, $\vec l_\bot$, is performed. 

Upon substituting (\ref{(5)}) and (\ref{(6)}) into 
(\ref{(18)}), and upon carrying out angular integrations, with 
(\ref{(8b)}), 
the diffractive forward production by transverse photons becomes,
\begin{eqnarray}
& & \frac{d\sigma_{\gamma^*_T p\rightarrow Xp}}{dt} 
\Bigg|_{t=0} = 
\frac{\alpha \cdot R_{e^+ e^-}}{3\cdot 16\pi^2} (\sigma^{(\infty)})^2 \int
\frac{dM^2}{M^2} \cdot \label{(20)}  \\
& & \cdot \left[ \frac{M^2}{Q^2 + M^2} - \frac{1}{2} \left( 1 + 
\frac{M^2 - \Lambda^2 (W^2) - Q^2}{\sqrt{(M^2 + \Lambda^2 (W^2) + Q^2)^2 - 4 
\Lambda^2 (W^2) M^2}} \right) \right]^2 , \nonumber
\end{eqnarray}
while for longitudinal ones, 
\begin{eqnarray}
& & \frac{d\sigma_{\gamma^*_L p\rightarrow Xp}}{dt} \Bigg|_{t=0} = 
\frac{\alpha \cdot R_{e^+ e^-}}{3\cdot 16\pi^2} (\sigma^{(\infty)})^2 \cdot 
 \label{(21)} \\
& &  \int \frac{dM^2}{Q^2} \left[ \frac{Q^2}{Q^2 + M^2} - 
\frac{Q^2}{\sqrt{(M^2 + \Lambda^2 (W^2) + 
Q^2)^2 - 4 \Lambda^2 (W^2)M^2}} \right]^2. \nonumber
\end{eqnarray} 
The integrands in (\ref{(20)}) and (\ref{(21)}) yield the mass spectra, 
$d\sigma_{\gamma^*_{T,L} p\rightarrow X p} / dt dM^2$,
for diffractive forward production $(t\cong 0)$ of 
states $X$
of unit spin by transversely and longitudinally polarized photons, 
respectively. 

Taking the sum of (\ref{(20)}) and (\ref{(21)}), 
one finds      
\begin{eqnarray}
& & \frac{d\sigma_{\gamma^* p \rightarrow Xp}}{dt} \Bigg|_{t=0} =   
\frac{\alpha R_{e^+ e^-}}{3\cdot 32\pi^2}(\sigma^{(\infty)})^2 \cdot
\label{(22)} \\
& & \cdot \int 
\frac{dM^2}{M^2} \left[ 1 - 
\frac{(M^2 + Q^2)^2 - (M^2 - Q^2) \Lambda^2 (W^2)}{(M^2 + Q^2) 
\sqrt{(M^2 + \Lambda^2 (W^2) + Q^2)^2 - 4 \Lambda^2 (W^2) M^2}} 
\right] . \nonumber
\end{eqnarray}
The integrand in (\ref{(22)}) yields the mass spectrum for the case of 
unpolarized photons.
Carrying out the integral in (\ref{(22)}), we find the total elastic forward 
production cross section in the mass interval $(M^2_1 , M^2_2)$,
\be
\frac{d\sigma_{\gamma^* p \rightarrow X p}^{(M^2_1 , M^2_2}}{dt} 
\Bigg|_{t=0} = \frac{\alpha
R_{e^+ e^-}}{3 \pi \cdot 16 \pi} \cdot \Pi_0 (Q^2, \Lambda^2 (W^2), M^2)
\Bigg|^{M^2_2}_{M^2_1}
\label{(23)}
\ee
with
\begin{eqnarray}
& & \Pi_0 (Q^2 , \Lambda^2 (W^2), M^2) =  
\frac{1}{2} \ln \frac{(\Lambda^2 + Q^2) (\sqrt X + Q^2 + \Lambda^2) + M^2
(Q^2 - \Lambda^2)}{\sqrt X + M^2 + Q^2 - \Lambda^2} - \nonumber \\
& & - \frac{\Lambda^2}{\sqrt{\Lambda^2 (4Q^2 + \Lambda^2)}} \ln 
\frac{\sqrt{\Lambda^2 (4Q^2 + \Lambda^2)} \sqrt X + \Lambda^2 (3 Q^2 - M^2 + 
\Lambda^2)}{Q^2 + M^2} , \label{(23a)}
\end{eqnarray}
where the short-hand
\be
X (M^2, Q^2, \Lambda^2 (W^2)) \equiv  (M^2 + \Lambda^2 (W^2) + Q^2)^2 - 4 
\Lambda^2 (W^2) M^2
\label{(23b)}
\ee
is being used and $\Lambda^2 \equiv \Lambda^2 (W^2)$ in (\ref{(23a)}). 

\section{The sum rules}

A comparison of the mass spectra in (\ref{(20)}) and (\ref{(21)}) with the 
expressions for the total cross section, $\sigma_{\gamma^*_{T,L}p}$,  
in (\ref{(11a)}) and (\ref{(12a)}) allows one to represent the 
total cross sections in 
terms of diffractive forward production, $\gamma^* p \rightarrow Xp$,
of states $X$ that carry photon quantum numbers (elastic diffraction). 
One finds the sum rules
\be
\sigma_{\gamma^*_T p} (W^2, Q^2) = \sqrt{16\pi} 
\sqrt{\frac{\alpha R_{e^+e^-}}{3 \pi}} \int_{m^2_0} dM^2 
\frac{M}{Q^2 + M^2} \sqrt{\frac{d\sigma_{\gamma^*_T}}{dt dM^2} \Bigg|_{t=0}}
\label{(24a)}
\ee
and  
\be
\sigma_{\gamma^*_L p}(W^2 , Q^2) = \sqrt{16\pi} 
\sqrt{\frac{\alpha R_{e^+e^-}}{3 \pi}} \int_{m^2_0} dM^2 
\frac{\sqrt{Q^2}}{Q^2 + M^2} \sqrt{\frac{d\sigma_{\gamma^*_L}}{dt dM^2} 
\Bigg|_{t=0}}.
\label{(25a)}
\ee
For the unpolarized cross section, by taking the sum of (\ref{(24a)}) and
(\ref{(25a)}), we have
\begin{eqnarray}
& & \sigma_{\gamma^* p} (W^2, Q^2) = \sqrt{16\pi} 
\sqrt{\frac{\alpha R_{e^+e^-}}{3 \pi}} \label{(26a)} \\
& & \cdot \int_{m^2_0} dM^2 
\frac{M}{Q^2 + M^2} \left[\sqrt{\frac{d\sigma_{\gamma^*_T}}{dt dM^2} \Bigg|_{t=0}}
+ \sqrt{\frac{Q^2}{M^2}} \sqrt{\frac{d\sigma_{\gamma^*_L}}{dt dM^2} \Bigg|_{t=0}}
\,\,\right] .\nonumber
\end{eqnarray}
In order to simplify the notation in (\ref{(25a)}) and (\ref{(26a)}),
we have dropped the arguments $W^2$, $Q^2$ $t$ and $M^2$ the diffractive 
production cross sections $\d\sigma_{\gamma^*_{T,L} p} / dM^2 dt$ depend on.
We stress that the sum rules (\ref{(24a)}) to (\ref{(26a)}) follow from the
two-gluon exchange structure of QCD that is contained in the representation
of the cross sections (\ref{(2)}) and (\ref{(16)}) in conjunction with the
form (\ref{(3)}) of the color-dipole cross section. The 
$\delta$-function ansatz (\ref{(8b)}) for the gluon-momentum distribution,
also used in the above derivation of the sum rules, does not introduce
much loss of generality. 
It is suggested and supported by the empirical scaling in $\eta$;
the gluon transverse momentum is fixed to coincide
with its average or effective value determined by $\Lambda^2 (W^2)$ \cite{2}.

It is amusing to note that (\ref{(26a)}) is the GVD analogue of the 
photoproduction sum rule from vector meson dominance \cite{7a}
\be
\sigma_{\gamma p}(W^2) = \sum_{V= \rho^0 , \omega , \phi, ...} 
\sqrt{16\pi} \sqrt{\frac{\alpha\pi}{\gamma^2_V}}
\sqrt{\frac{d\sigma_{\gamma p\rightarrow V_0}}{dt} \Bigg|_{t=0}}.
\label{(27a)}
\ee
Indeed, multiplying the (imaginary part of the) amplitude for $\gamma^*p
\rightarrow X p$ by the propagator factor $M^2/(Q^2 + M^2)$ and a
factor $1/M$ for the strength of the photon coupling normalized by
$\sqrt{\alpha \cdot R_{e^+e^-}/3\pi}$, implies (\ref{(24a)}).\footnote{Based
on this GVD argument, the sum rule (\ref{(24a)}) was indeed given 
before\cite{9}.} An additional well-known factor of 
$\sqrt{Q^2/M^2}$\cite{Fraas} is needed for the longitudinal cross section
in (\ref{(25a)}).
Note that this derivation of the sum rules (\ref{(24a)}) to (\ref{(26a)}) is 
dependent on the production mechanism for $\gamma^* p \rightarrow Xp$ in so far 
only, as the $Q^2$ dependence induced by the transition from the timelike
four momentum squared, $P_X^2=M^2$, of the final state $X$ to the 
spacelike four momentum squared, $Q^2$, of the virtual photon coupled
to $X$ is assumed to be fully contained in the aforementioned
(propagator) factors.  In other words, it is assumed that the underlying 
transition from the timelike to spacelike four momenta with respect to 
the vector state $X$ does not affect $\d\sigma_{\gamma^*_{T,L} p} / dM^2 dt$
at $t=0$.
This condition is fulfilled in the GVD/CDP based on (\ref{(2)}),
(\ref{(3)}) and (\ref{(16)}) with (\ref{(8b)}).

The above derivation of the sum rules (\ref{(24a)}) to (\ref{(26a)}), based
on the comparison of the QCD-based mass spectra for elastic diffraction in 
(\ref{(20)}) and (\ref{(21)}) with the ones for $\sigma_{\gamma^*_{T,L} p}$
in (\ref{(11a)}) and (\ref{(12a)}), demonstrates explicitly that the QCD-based
color-dipole picture implies
GVD for low-x DIS. The terminology GVD/CDP for our approach to DIS at
low x is appropriate.

The experimental validity of the photoproduction sum rule (\ref{(27a)}) 
was carefully
investigated in the late sixties and the early seventies\cite{Wolf}. Insertion
of the experimental data for the total photoproduction cross section on the
left-hand side in (\ref{(27a)}), and of the cross sections for vector meson
forward production on the right-hand side, revealed a discrepancy of 22 \%
that led to the formulation of generalized vector dominance\cite{Sakurai}. 
A recent experimental test of (\ref{(27a)}) at HERA energies was
presented in ref. \cite{ZEUS}.

An analogous direct experimental test of the sum rules (\ref{(24a)}) to
(\ref{(26a)}) for virtual photons is more difficult to be carried out. The
diffractive forward production cross sections in (\ref{(16)}) and
(\ref{(24a)}) to (\ref{(26a)}) refer to the production of spin 1 (vector)
states; otherwise the produced states would never couple to the photon
and build up the imaginary part of the
(virtual) forward Compton scattering amplitude. A direct experimental
verification of (\ref{(24a)}) to (\ref{(26a)}), accordingly, requires the
projection of the spin 1 component contained
in the diffractively produced state $X$
of mass $M$. In addition, diffractive production by transverse and
longitudinal  photons has to be separated; an assumption of 
s-channel helicity 
conservation\cite{Fraas}, as in vector meson production, may be helpful, 
as long as no
direct separation of production by transverse and longitudinal photons
will be available.

We note that, independently of their direct experimental verification, the sum 
rules (\ref{(24a)}) to (\ref{(26a)}) are of theoretical interest.
They most explicitly demonstrate that (\ref{(16)}) (containing a dipole cross
section, $\sigma_{(q \bar q)p}$, identical to the one in $\sigma_{\gamma^* p}$
in (\ref{(2)})) describes {\it elastic} diffraction, elastic with 
respect to the photon quantum numbers carried by the incoming and outgoing 
$q \bar q$ states; had the diffractively produced state $X$ quantum 
numbers different from the ones of the photon, e.g. a different spin, the 
sum rules (\ref{(24a)}) to (\ref{(26a)}) could never be valid. 
In order to incorporate {\it inelastic} diffraction, the dipole cross section 
in the expression for diffractive production (\ref{(16)}) has to be replaced 
by the addition of a term describing inelastic diffraction.

\section{Comparison with experimental results on diffractive production }  

\begin{figure}[htbp]\centering
\epsfysize=15cm
\centerline{\epsffile{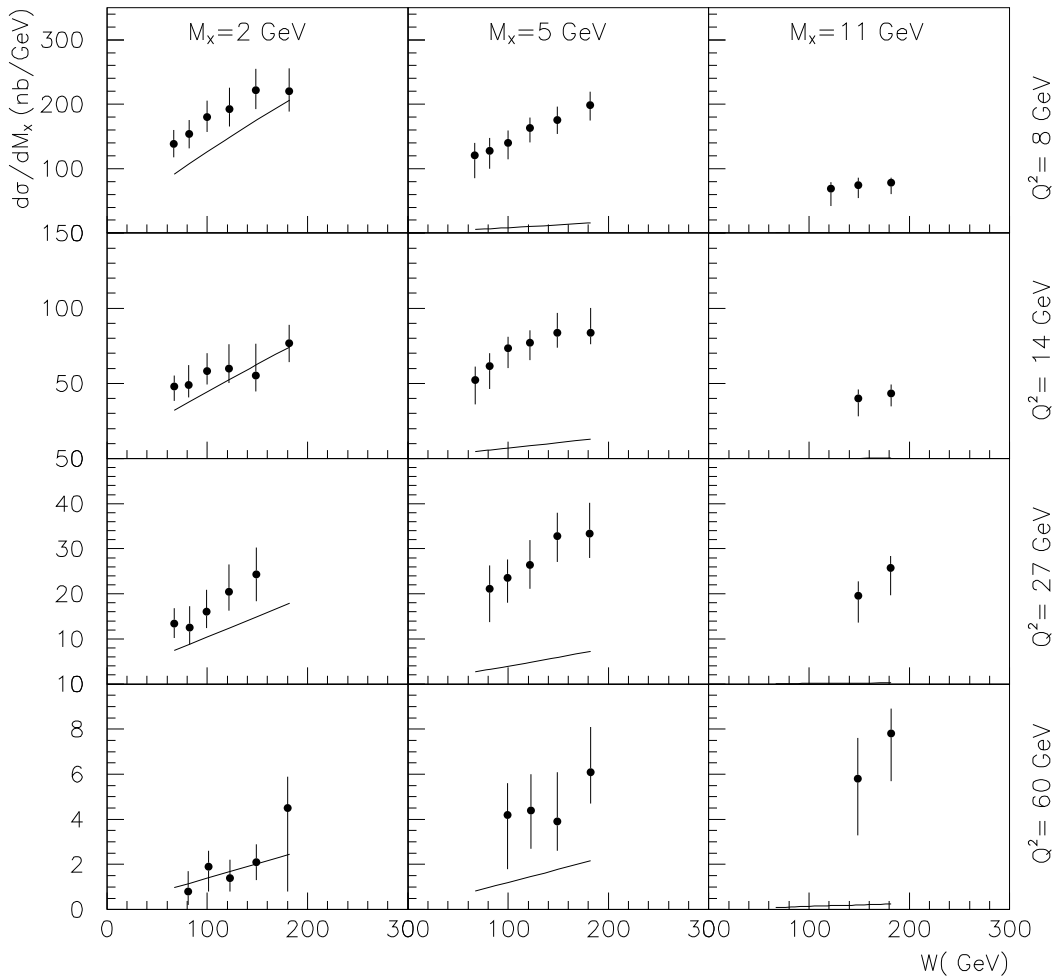}}
\caption{  The ZEUS data for diffractive production, $\gamma^*p
\rightarrow Xp$, as a function of the energy, $W$, for different
masses $M_X$ and photon virtuality $Q^2$ compared with the GVD/CDP 
predictions for elastic
diffraction. The excess of the data with respect to theory is due to 
diffractive
production of states $X$ of mass $M_X$ that do not couple to the photon,
and accordingly cannot contribute to the (virtual) forward Compton
amplitude that builds up the total cross section, $\sigma_{\gamma^*p}$.}

\label{Fig. 1}
\end{figure}

\begin{figure}[htbp]\centering
\epsfysize=18cm
\centerline{\epsffile{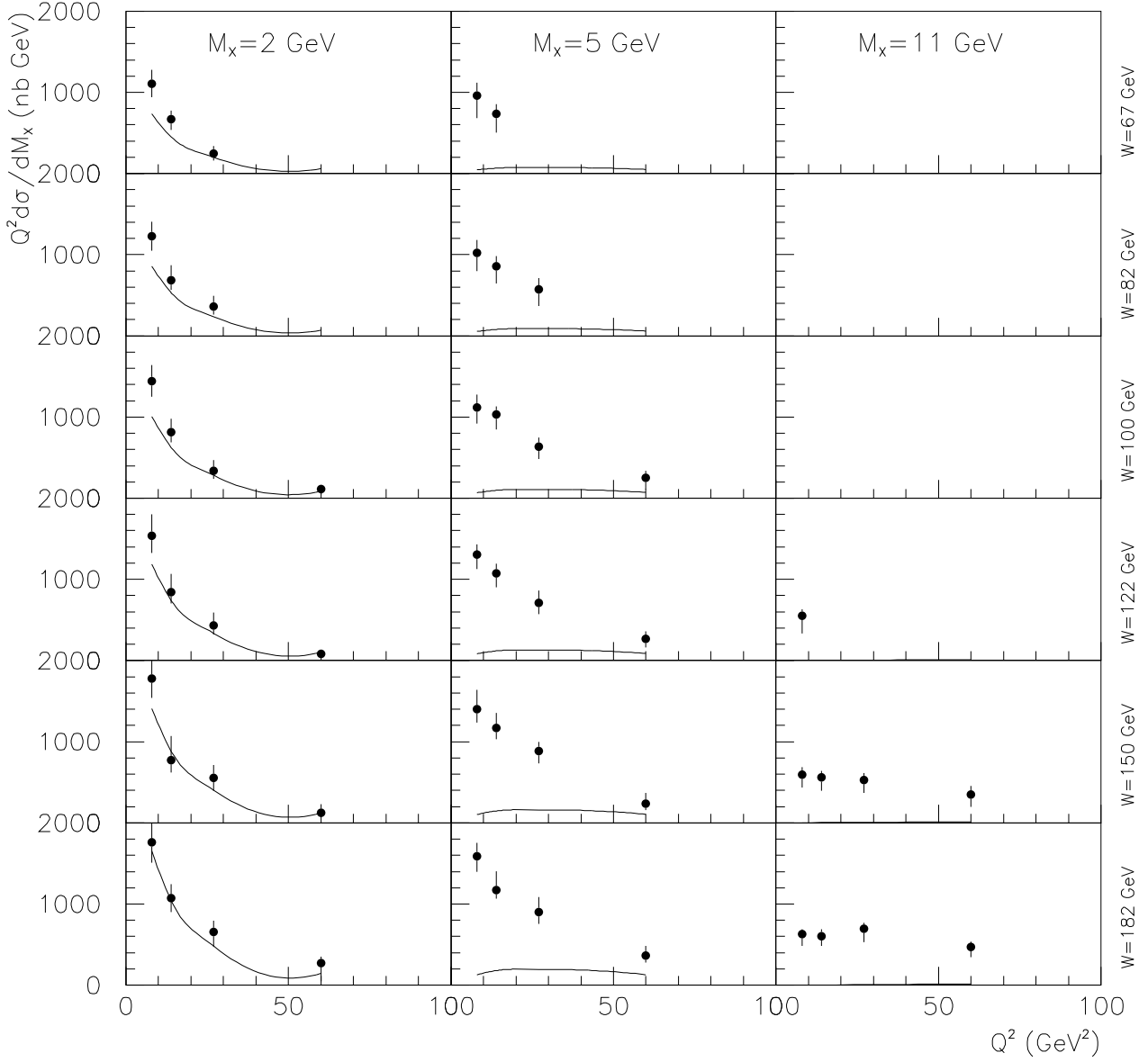}}
\caption{  As in fig. 1, but for $Q^2 d\sigma_{\gamma^* p \rightarrow
X_p}/dM_X$ as a function of $Q^2$ for different masses $M_X$.}
\label{Fig. 2}
\end{figure}

\begin{figure}[htbp]\centering
\epsfysize=14cm
\centerline{\epsffile{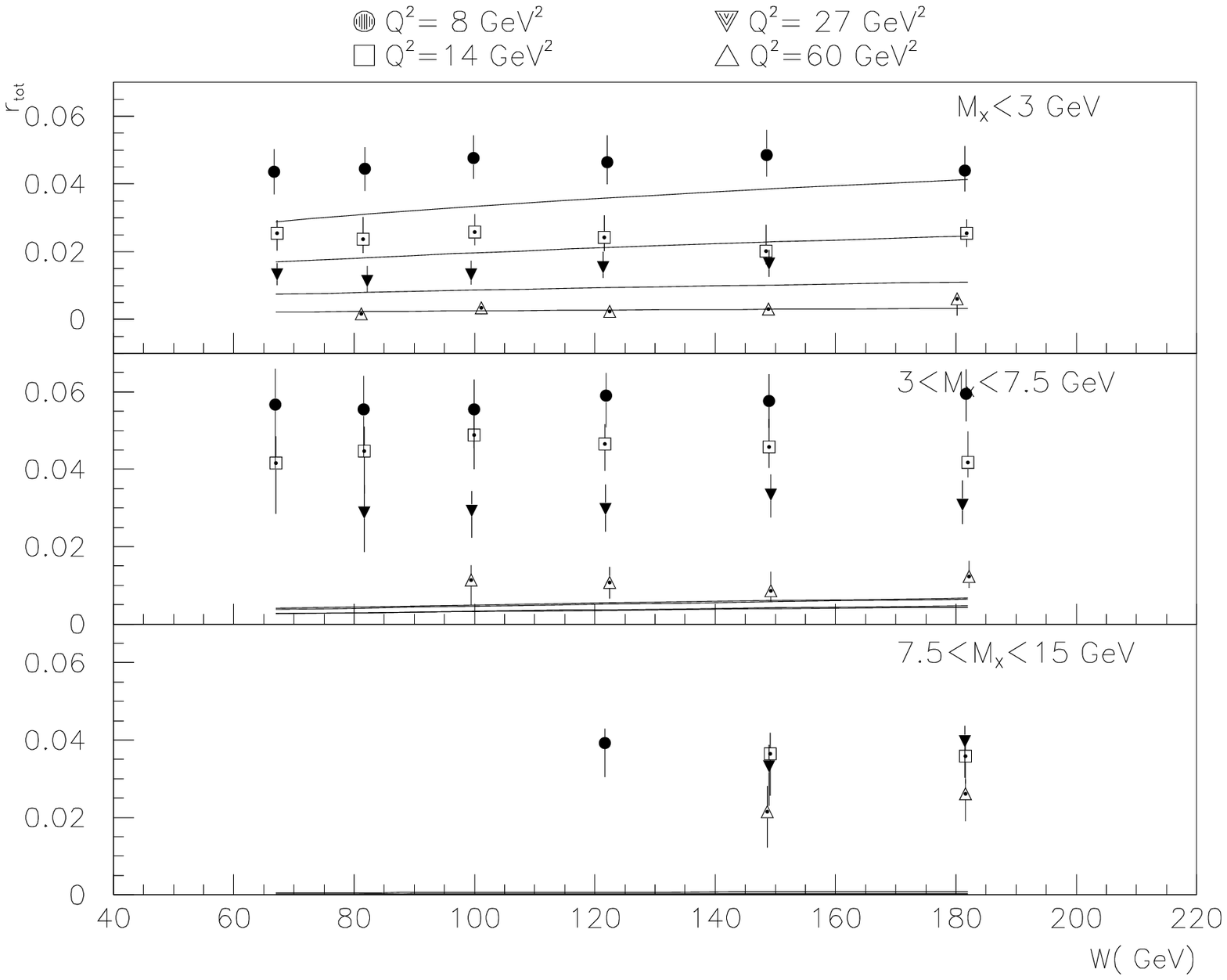}}
\caption{ The ratio of the cross section for diffractive production 
to the total cross section as a function of the energy, $W$, for different
values of $M_X$ and $Q^2$. The theoretical curves, 
as in figs. 1,2, show the predictions from
the GVD/CDP for the component of diffraction that saturates the imaginary
part of the virtual forward Compton amplitude (elastic diffraction).}

\label{Fig. 3}
\end{figure}

We turn to a comparison of our results on elastic diffractive production 
with the experimental data. The 
ZEUS collaboration has presented data \cite{8} for the mass distribution
integrated over the distribution in momentum transfer $t$. Assuming an 
exponential behavior, $\exp (-b t)$, the experimental data 
on $d\sigma_{\gamma^* p\rightarrow Xp}/dM_X$ are related to the 
mass spectrum in the integrand of (\ref{(22)}) by 
\be
\frac{d\sigma_{\gamma^* p\rightarrow Xp}}{dM_X} = 2 M_X \int dt~ e^{-bt} 
\frac{d\sigma_{\gamma^* p\rightarrow Xp}}{dtdM^2_X} \Bigg|_{t=0} 
= \frac{2 M_X}{b} 
\frac{d\sigma_{\gamma^* p\rightarrow Xp}}{dt dM^2_X} 
 \Bigg|_{t=0}.
\label{(24)}
\ee
In (\ref{(24)}), we use the notation of  
the ZEUS collaboration by 
the replacement $M \rightarrow M_X$. In fig. 1,
following the representation of the data by the ZEUS collaboration,
we show the energy dependence of $d\sigma_{\gamma^* p\rightarrow Xp} / dM_X$ 
for various masses $M_X$ and various fixed values of $Q^2 = 8 {\rm GeV}^2$ 
to $Q^2 = 60 {\rm GeV}^2$.
As in the total cross section (\ref{(11)}), in (\ref{(22)}) and
(\ref{(24)}), we have used\footnote{The total cross section is equally well
represented in the three-flavour option, $R_{e^+e^-} = 2$ but 
$\sigma^{(\infty)} = 80 {\rm GeV}^{-2}$. In this case, $b = 12.5 
{\rm GeV}^{-2}$ is to be used.} $R_{e^+e^-} = 10/3$ as well as
$\sigma^{(\infty)} = 48 {\rm GeV}^{-2}$, and\footnote{A detailed analysis of 
the effect of a potential $Q^2$ or $M^2$ dependence of the slope parameter
seems somewhat premature in view of the available data and the incompleteness
of the theory with respect to inelastic diffraction.}

 $b = 7.5 {\rm GeV}^{-2}$.
Figure 1 shows that diffractive production at low masses is approximately
described by elastic diffraction, i.e. by diffractive production of
those spin 1 states that saturate the (imaginary part of the virtual)
Compton forward scattering amplitude. For the higher masses of
$M_X = 5 {\rm GeV}$ and $M_X = 11 {\rm GeV}$, as expected, elastic diffraction
does by no means fully represent the diffractive cross section. The 
discrepancy between theory and experiment decreases with increasing $Q^2$,
however, i.e. with increasing $\beta \equiv Q^2/(Q^2 + M^2_X)$. 
In fig. 2, we show the $Q^2$
dependence in a plot of $Q^2 d\sigma_{\gamma^* p \rightarrow X_p}/dM_X$
against $Q^2$. As anticipated from fig. 1, for $M_X = 2 GeV$ there is 
reasonable consistency between the theoretically calculated production
of spin 1 states and the experimental data.

In fig. 3, we show the ratio
\be
r_{{\rm tot}} = \frac{\int^{M_b}_{M_a} dM_X d\sigma_{\gamma^* p\rightarrow Xp}
/dM_X}{\sigma_{\gamma^* p}}\label{(35a)}
\ee
as a function of the energy $W$. As anticipated from the previous 
figures, there is some discrepancy in normalization and also in energy 
dependence.
Note, however, that the energy dependence, as a consequence of the 
different structure of the expressions for $\sigma_{\gamma^* p}$ in 
(\ref{(7)}) and for  
$d\sigma_{\gamma^* p\rightarrow Xp}/dt$ in (\ref{(18)}), is quite similar. 
The naive expectation (\ref{(11)}) that the linearity 
with respect to the dipole cross section in the total cross section (\ref{(2)})
and the non-linearity in the diffractive production cross section 
(\ref{(16)}) lead to distinctly different energy dependences is not valid. 

The substantial excess of the cross section 
$d\sigma_{\gamma^*p \rightarrow Xp}/dM_X$
for $M_X = 5 {\rm GeV}$ and $M_X = 11 {\rm GeV}$ is due to the production
of states that do not couple to the photon and accordingly do not 
contribute to the imaginary part of the virtual forward Compton scattering
amplitude (inelastic diffraction). From the thrust and sphericity 
analysis of the diffractively produced states \cite{ZEUS2}, we know that these
predominantly consist of hadronized $q \bar q$ states with some (fairly
small) admixture of a $q \bar q$ + gluon component. The excess in
$d\sigma_{\gamma^*p \rightarrow Xp}/dM_X$ 
must be associated with states of spin higher than the
photon spin. With increasing $Q^2$, due to propagator effects,
the relative contribution of low masses becomes increasingly more
suppressed, the diffraction process becomes more elastic. The discrepancy
between our theoretical curves and the experimental data is decreased
at high $Q^2$, or, in terms of the frequently employed variable $\beta = Q^2 /
(Q^2 + M^2_X)$, elastic diffraction becomes dominant for $\beta \rightarrow 1$.

Any theory of high-mass (small $\beta$) diffraction has to 
obey the constraint that the
component needed in addition to the elastic one be truly 
inelastic in the sense of
being unable to contribute to the imaginary part of the forward Compton
scattering
amplitude. After all, the forward Compton amplitude is saturated by the
elastic component which contributes the amount to diffractive production
that is shown in figs. 1 to 3. It is worth stressing again the fairly general
validity of the sum rules (\ref{(24a)}) to (\ref{(26a)}), this saturation 
property is based on.

Some recent theoretical work \cite{Golec,Forshaw,Bartels} 
on diffractive production of large masses, i.e. $\beta \ll 1$, concentrated 
on adding a 
quark-antiquark-gluon $(q \bar q g)$ component to the $q \bar q$ wave function of 
the photon. Even though the data on diffractive production are accounted for, 
this approach suffers from a serious inconsistency, as, without 
justification, the 
additional $q \bar q g$ component is only taken into account in 
the photon wave function entering diffractive production (i.e. in (\ref{(16)})), 
while being ignored in the total cross section (i.e. in (\ref{(2)})), thus
disregarding the optical theorem. 
From a theoretical point of view, the interplay of $q \bar q$ and $q \bar q g$ 
components in the wave function for diffractive production and the total cross 
section is analyzed in \cite{10}. 
The approach of ref.\cite{10} treats the $q \bar q$ and $q \bar q g$ 
components of the photon on 
equal footing. It does not contain a truly inelastic component. Accepting 
a universal (model-independent) validity of the sum rules in Section 4, an 
approach purely based on an elastic component that describes diffractive 
production is likely to fail for the total cross section. Indeed, inserting an 
amplitude for diffractive production into the sum rules that coincides 
with experiment,
the resulting total cross section is likely to disagreee with 
experiment, since agreement with experiment is achieved by the much smaller 
amplitude for elastic diffractive production by itself. 
A clear discrimination between elastic and 
inelastic diffractive production is made right from the outset in the 
color-dipole approach of \cite{Bialas}. 
It is not entirely clear from the presentation in \cite{Bialas} by what 
means a potential contribution of the ``inelastic'' component to the 
imaginary part of the virtual forward Compton scattering amplitude is 
excluded.

\section{Implications for $\sigma_{\gamma^* p}$}

In view of the comparison of our results for elastic diffraction with the
experimental data, it will be enlightening to return to the theoretical
description of the total cross section, $\sigma_{\gamma^*p}$. The strong
decrease of the theoretical results for elastic
diffractive production with increasing mass, $M \equiv M_X$,
by no means implies that contributions due to large masses in the integral
representations (\ref{(11a)}) and (\ref{(12a)}), or, equivalently, 
(\ref{(24a)}) and (\ref{(25a)}), for the transverse and
longitudinal total cross sections
become negligible. This may be explicitly seen by evaluating the integral
representations as a function of an upper limit,
$m^2_1$.
Indeed, the sum rules (\ref{(24a)}) and (\ref{(25a)}) suggest this upper limit
to be approximately given by the upper end of the diffractively produced 
spectrum of masses. 

\begin{figure}[htbp]\centering
\setlength{\unitlength}{1cm}
\begin{minipage}[t]{8.0cm}
\vspace*{2.4cm} 
\begin{picture}(3.5,3.5)\psfig{file=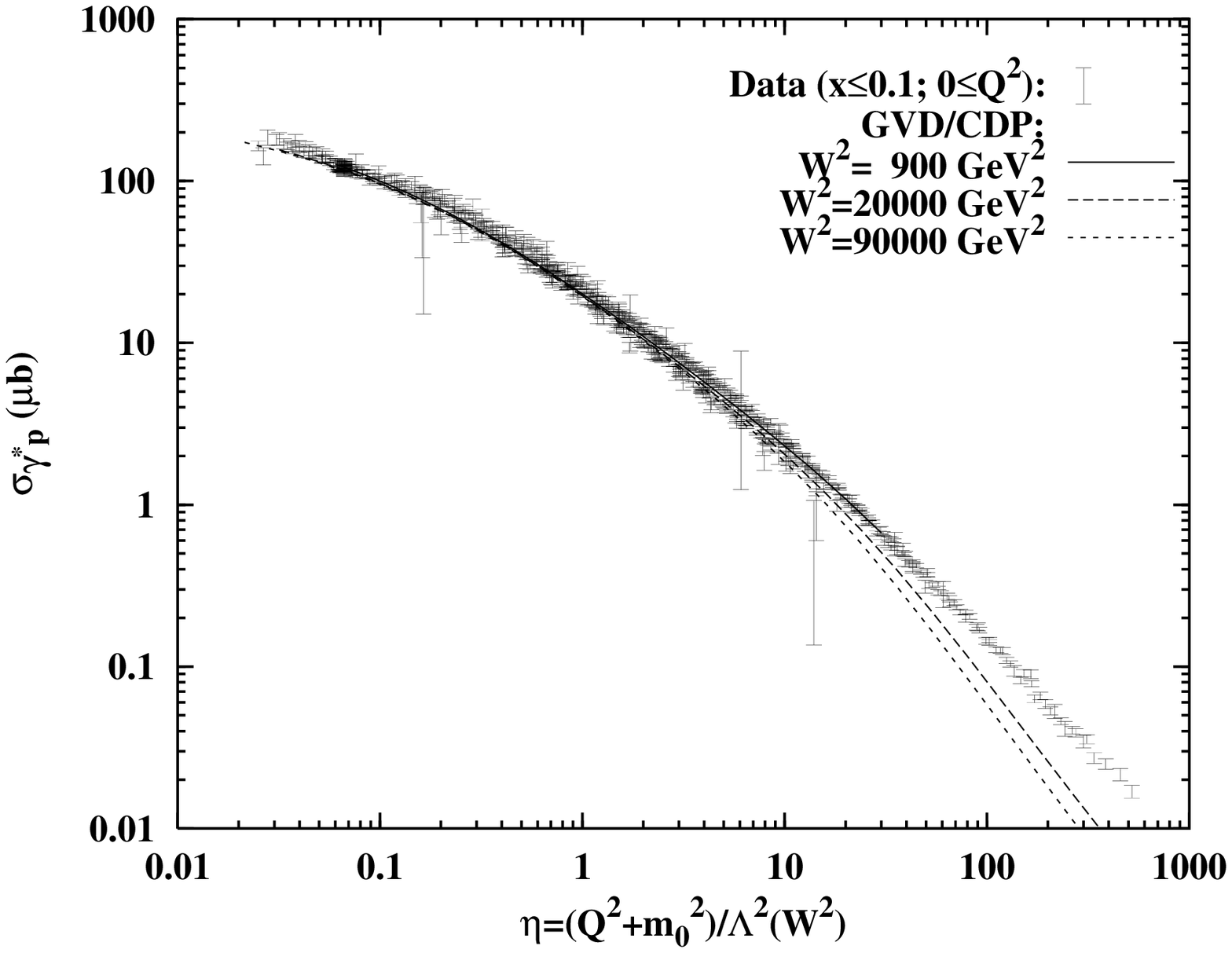,width=8.0cm}
\put(-6,3){\large a)}
\end{picture}\par
\end{minipage}\hfill

\begin{minipage}[t]{8.0cm}
\vspace*{2.4cm}
\begin{picture}(3.5,3.5)\psfig{file=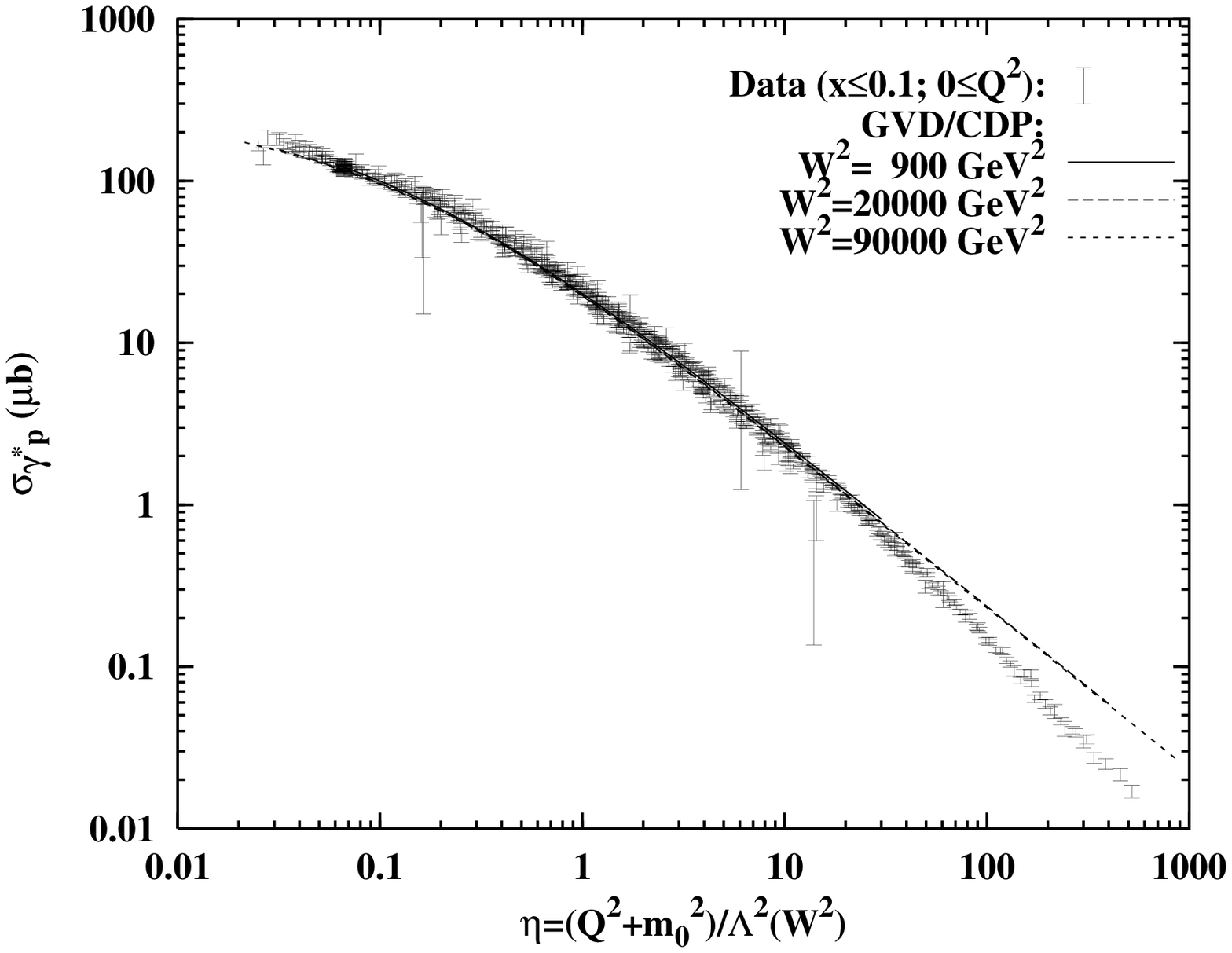,width=8.0cm}
\put(-6,3){\large b)}
\end{picture}\par
\end{minipage}\hfill

\setlength{\unitlength}{1cm}
\begin{minipage}[t]{8.0cm}
\vspace*{2.4cm}
\begin{picture}(3.5,3.5)\psfig{file=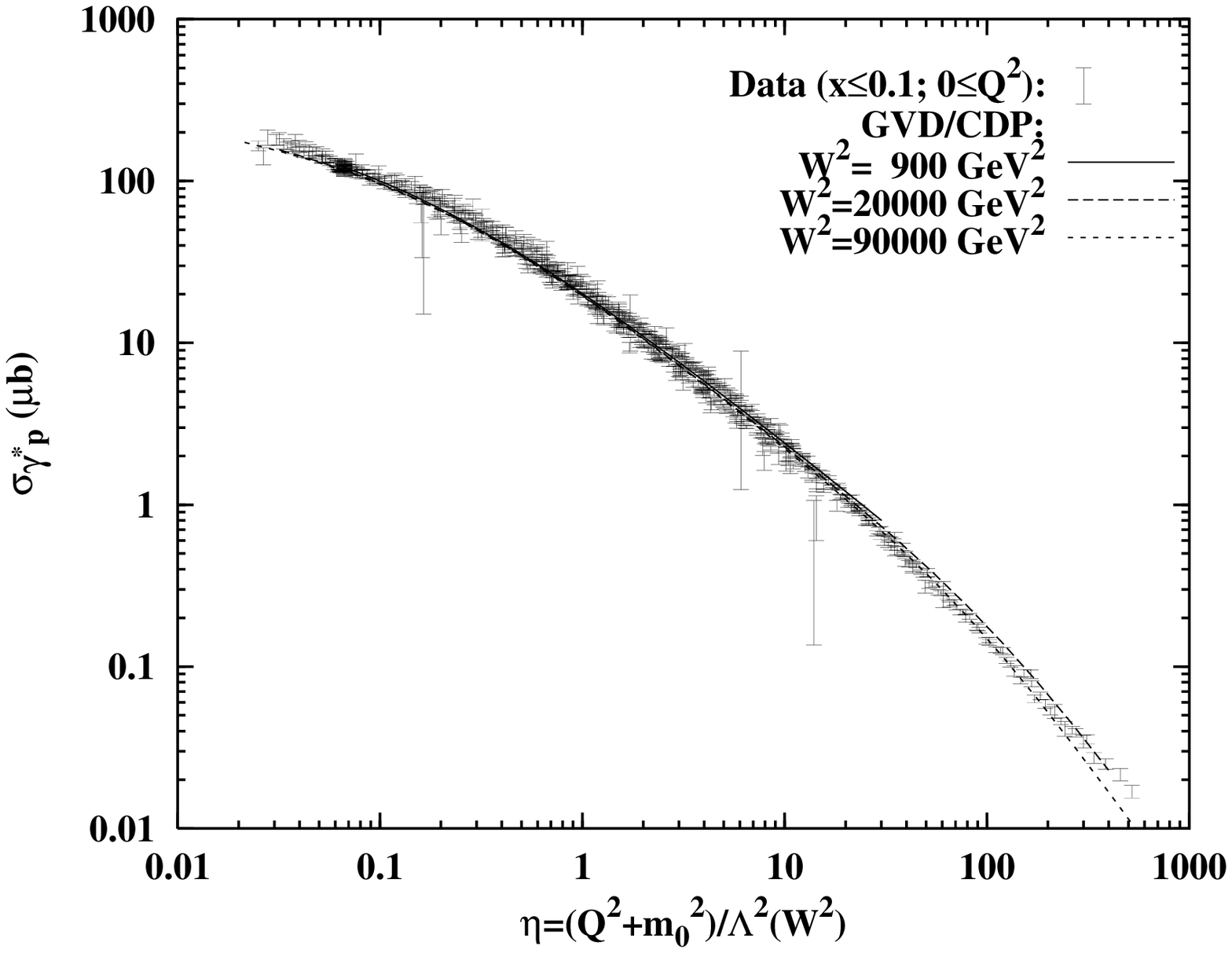,width=8.0cm}
\put(-6,3){\large c)}
\end{picture}\par
\end{minipage}\hfill
\caption{   The data for the total photoabsorption cross section,
$\sigma_{\gamma^*p}$, as a 
function of the scaling variable $\eta$ compared with the GVD/CDP predictions
obtained for different values of the upper bound, $m^2_1$, on the diffractive
mass spectrum that is integrated over in the imaginary part of the virtual
forward Compton scattering amplitude describing $\sigma_{\gamma^*p}$.
Figures 4a to 4c refer to $m^2_1 = 100 \,{\rm GeV}^2$, $m^2_1 = \infty$ and 
$m^2_1 = 484 \, {\rm GeV}^2$, respectively.}
\label{Fig. 4}
\end{figure}

The integration of (\ref{(11a)}) and (\ref{(12a)}) then yields
\begin{eqnarray}
& & \sigma_{\gamma^*_Tp} = \frac{\alpha R_{e^+e^-}}{3 \pi} \sigma^{(\infty)}
\Bigg[ \frac{Q^2}{Q^2+M^2} + \frac{1}{2} ln \frac{Q^2 + M^2}{\sqrt X + Q^2
+ M^2 - \Lambda^2} \label{(36)} \\
& & - \frac{2 Q^2 + \Lambda^2}{2 \sqrt{\Lambda^2 (\Lambda^2
+ 4Q^2)}} ln \frac{\sqrt{\Lambda^2 (\Lambda^2 + 4 Q^2) X} + \Lambda^2 (3 Q^2
- M^2 + \Lambda^2)}{Q^2 + M^2}\Bigg] \Bigg|^{m^2_1}_{m^2_0} \nonumber
\end{eqnarray}
and
\begin{eqnarray}
& & \sigma_{\gamma^*_L p} = \frac{\alpha R_{e^+e^-}}{3 \pi} \sigma^{(\infty)}
\Bigg[ \frac{-Q^2}{Q^2 + M^2} \label{(37)} \\
& & + \frac{Q^2}{\sqrt{\Lambda^2 (\Lambda^2 + 4Q^2)}}
ln \frac{\sqrt{\Lambda^2 (\Lambda^2 + 4 Q^2) X} + \Lambda^2 (3 Q^2
- M^2 + \Lambda^2)}{Q^2 + M^2}\Bigg] \Bigg|^{m^2_1}_{m^2_0}, \nonumber
\end{eqnarray}
where $\Lambda^2 (W^2)$ is given in (\ref{(9)}), and $X (M^2 , Q^2, \Lambda^2 
(W^2))$ is defined in (\ref{(23b)}).
The sum of (\ref{(36)}) and (\ref{(37)}), for $m^2_1 = \infty$, reduces
to (\ref{(11b)}). For completeness, we also give the 
previously mentioned correction terms that
have to be added to (\ref{(11a)}) and (\ref{(12a)}) as well as to 
(\ref{(36)}) and (\ref{(37)}), in order to assure
identical lower limits, $m^2_0$, in the initial and the final state of the
(virtual) Compton forward scattering amplitude. The expressions given
for these terms in ref. \cite{1}
may be simplified to become one-dimensional integrals that are to be carried
out numerically. For the realistic case of $\Lambda^2 > 4 m^2_0$, one
finds,
\begin{eqnarray}
\Delta \sigma_{\gamma^*_T p} & = &\frac{\alpha R_{e^+e^-}}{6 \pi} 
\sigma^{(\infty)} \int_{(\Lambda - m_0)^2}^{(\Lambda + m_0)^2} dM^2 
\frac{1}{(Q^2 + M^2)} \label{(38)} \\
& & \Bigg[ \frac{1}{2 \pi} \arccos \left( 
\frac{\Lambda^2 + M^2 - m^2_0}{2 M \Lambda} \right) +
\frac{M^2 - \Lambda^2 - Q^2}{\pi \sqrt{X}} \arctan \sqrt{Y} \Bigg],
\nonumber
\end{eqnarray}
and
\be
\Delta \sigma_{\gamma^*_L p} = \frac{\alpha R_{e^+e^-}}{6 \pi} 
\sigma^{(\infty)} \int_{(\Lambda - m_0)^2}^{(\Lambda + m_0)^2} dM^2 
\frac{1}{(Q^2 + M^2)} \cdot  \frac{2}{\pi \sqrt{X}} \arctan \sqrt{Y},
\label{(39)}
\ee
where
\be
Y \equiv \frac{(M + \Lambda)^2 + Q^2}{(M-\Lambda)^2 + Q^2} \cdot
\frac{m^2_0 - (M - \Lambda)^2}{(M + \Lambda)^2 - m^2_0}.\label{(39a)}
\ee

A comparison of the theoretical results 
for $\sigma_{\gamma^*p}$ for different values of the upper
limit, $m^2_1$, with the experimental data \cite{23,24,25,26,27}
is shown in fig. 4. We emphasize that the experimental
data cover the full range of the kinematic variables $(x < 0.1$, all $Q^2$, 
including $Q^2 = 0$)
where scaling in $\eta$ was established in a 
model-independent analysis\cite{1}. The results in fig. 4 are quite remarkable.
They show that 

\begin{itemize}
\item[i)] a restriction of $m^2_1$ to values (e.g. $m^2_1 = 100 \, {\rm GeV}^2$)
below the upper limit of the mass
where appreciable diffractive production was observed experimentally, 
leads to values of
$\sigma_{\gamma^*p}$ that for $\eta \ge 10$ lie much below the 
experimental scaling curve, i.e. a non-vanishing elastic diffractive production 
of large masses $M_X$
is necessary for saturation of the forward Compton amplitude, even though
\item[ii)] the previously used \cite{1, 2}
value of $m^2_1 = \infty$, for $\eta \gsim 10$ leads to results that
lie above the experimental data.
\end{itemize}
The highest mass bin where appreciable diffractive production occurs, 
according to the results \cite{ZEUS2}
from the ZEUS collaboration, is given by $M_X = 22$ GeV. Accordingly we use
\be
m^2_1 \simeq (22 \,{\rm GeV})^2 = 484 \, {\rm GeV}^2
\label{(40)}
\ee
that yields good agreement with the experimental data for $\sigma_{\gamma^* p}
(\eta)$ in the full kinematic range of $x \le 0.1$,
all $Q^2$ including $Q^2 = 0$, where scaling in $\eta$ was established 
in a model-independent analysis of the experimental data.

The fact that an upper limit for the diffractively produced mass should enter 
the forward Compton amplitude at finite energy, $W$, is not unexpected. It
is gratifying that its value (\ref{(40)}) of $m_1 = 22 \,{\rm GeV}$,
coincides with the upper limit of the range
of masses where diffractive production, $\gamma^* p \rightarrow Xp$, is
experimentally found to occur. Beyond that mass, inclusion of the mass 
spectrum from 
the GVD/CDP overestimates the total cross section, $\sigma_{\gamma^* p}$.
This deficiency, in an approximate and admittedly somewhat crude manner,
is repaired by restricting the range of integration by the upper limit
in (\ref{(40)}). 
The deviation from scaling in $\eta$ resulting from the introduction of this 
upper limit is a fairly mild one.

\section{Conclusion}

We summarize as follows:
\begin{itemize}
\item[i)]
Our ansatz for DIS at low $x$ that is based on the generic structure of 
two-gluon exchange from QCD supplemented by the empirical scaling behavior 
$\sigma_{\gamma^* p} = \sigma_{\gamma^* p} (\eta)$ leads to sum rules that 
explicitly express $\sigma_{\gamma^* p}$ as an appropriate integral over the 
square root of the elastic diffractive forward production cross section. The 
agreement of the expression for $\sigma_{\gamma^* p}$ with the experimental 
data explicitly demonstrates that DIS at low $x$ is understood in terms of 
diffractive forward scattering of massive $q \bar q$ pairs on the nucleon
(GVD/CDP). 

\item[ii)]
A comparison of the theoretical results for elastic diffractive production, 
i.e. the production of states $X$ in $\gamma^* p \rightarrow Xp$ that carry
photon quantum numbers, is the dominant mechanism for $\beta \equiv 
Q^2 / (Q^2 + M^2_X) \rightarrow 1$. 
The excess of the experimental data with respect to elastic diffraction 
observed for $\beta \ll 1$ is 
attributed to inelastic diffraction, i.e. to the production of states $X$ that 
do not carry photon quantum numbers and are not contributing to the imaginary part
of the virtual Compton forward scattering amplitude. 

\item[iii)]
The connection between the total photoabsorption cross section and 
diffractive production suggests an extension of the kinematic range, in which
the total photoabsorption cross section, $\sigma_{\gamma^* p}$, is 
represented by the GVD/CDP. This range now includes the full kinematic 
region where scaling in $\eta$ holds, 
$x \le 0.1$ and all $Q^2 \ge 0$.  

\item[iv)]
Any serious theoretical model for diffractive production must be 
examined with respect to its compatability with the experimental data for the 
total virtual photoabsorption cross section that is related to (elastic)
diffraction via the optical theorem. 
\end{itemize}

\vspace{0.5cm}
{\Large\bf Acknowledgement}    

\bigskip
One of the authors (D.S.) thanks the MPI in Munich, where this work was
finalized, for warm hospitality and C.~Kiesling and L.~Stodolsky for
useful discussions.
The support by M. Tentyukov in the presentation of the data is 
gratefully acknowledged.

\end{document}